\documentclass[12pt]{article}
\usepackage[dvips]{graphics,color}
\usepackage{aasms4}
\usepackage{epsfig}

\begin{document}
\title{\bf CONSTRAINING COSMOLOGICAL MODELS BY THE CLUSTER MASS FUNCTION}
 
\author{Nurur Rahman \ and \ Sergei F. Shandarin 
\\ Department of Physics and Astronomy, \ University of Kansas, \ Lawrence, KS 66045; 
\\ nurur@kusmos.phsx.ukans.edu, \ sergei@ukans.edu}

\date{Janu12th 2000}
\begin{abstract}
We present a comparison between two observational and three theoretical 
mass functions for eight cosmological models suggested by the data 
from the recently completed BOOMERANG-98 and MAXIMA-1 cosmic microwave 
background (CMB) anisotropy experiments as well as peculiar velocities (PVs) 
and type Ia supernovae (SN) observations. The cosmological models have been 
proposed as the best fit models by several groups. 
We show that no model is in agreement with the abundances of X-ray clusters
at $\sim 10^{14.7} h^{-1}M_{\odot}$.
On the other hand, we find that the BOOM+MAX+{\sl COBE}:I, Refined Concordance
and $\Lambda$MDM are in a  good agreement with the abundances of optical
clusters. The P11 and especially Concordance models predict a slightly lower
abundances than observed at $\sim 10^{14.6} h^{-1}M_{\odot}$. The 
BOOM+MAX+{\sl COBE}:II and PV+CMB+SN models predict a slightly higher
abundances than observed at $\sim 10^{14.9} h^{-1}M_{\odot}$.
The nonflat MAXIMA-1 is in a fatal conflict with the observational cluster 
abundances and can be safely ruled out.
\end{abstract}

\keywords{cosmology:theory --- cosmology:observation --- 
galaxies: clusters: general --- large-scale structure of universe}

\section{Introduction}
Recently, certain cosmological models have received a fairly strong 
observational boost. 
Several groups have used the new cosmic microwave background (CMB) data from 
the BOOMERANG-98 (the 1998 Balloon Observations Of Millimetric Extragalatic 
Radiation ANd Geophysics; \cite{deb-etal00}) and 
MAXIMA-1 (The first overnight flight of the Millimeter Anisotropy eXperiment
IMaging Array; \cite{han-etal00}) anisotropy experiments to constrain 
cosmological parameters. Other groups combined the  constraints from CMB with 
cosmological nucleosynthesis data, peculiar velocities (PVs) and type Ia 
supernovae (SN) observations. The values of cosmological parameters vary 
from one set to the next, but all of these models are in reasonable agreement 
with a flat Cold Dark Model (CDM) universe
($\Omega_{0} + \Omega_{\Lambda} = 1$) dominated by the vacuum energy
except MAXIMA-1 with matter density, $\Omega_0=0.68$ and vacuum energy
density, $\Omega_{\Lambda}=0.23$ (Balbi et al. 2000). 

In this Letter, we compare the abundances of clusters of galaxies
predicted by some popular cosmological models with observed abundances.
The abundance of clusters has been shown to be one of the simplest but most 
effective cosmological tools for constraining the models of structure 
formation.
It can place strong constraints on the parameters  of cosmological models
(\cite{kai86}, Peebles, Daly, \& Juskiewicz 1989, \cite{sim-sh89}), 
including the mass density in the universe ($\Omega_{0}$) 
and the amplitude of the mass density fluctuations ($\sigma_{8}$) or, 
equivalently, the bias factor ($b=1/\sigma_{8}$; \cite{evr89}; 
\cite{fre-etal90}; \cite{hen-arn91}; \cite{bah-cen92}; \cite{lil92}; 
\cite{ouk-bla92}; Kofman, Gnedin, \& Bahcall 1993; White, Efsthatiou, \& 
Frenk 1993; \cite{bon-mye96}; Eke, Cole, \& Frenk 1996; Mo, Jing, \& White 
1996; \cite{via-lid96}; \cite{bor-etal97};  \cite{hen97}; \cite{pen98}; 
\cite{pos99}; \cite{ver-etal01}; Pierpaoli, Scott, \& White 2001).
  
The abundance of clusters and their evolution are quantified by the mass 
distribution function. The theoretical derivation of the mass function
of gravitationally bound objects has been pioneered by 
Press \& Schechter (1974, hereafter PS). Despite various modifications 
that have been suggested recently (Cavaliere, Colafrancesco, \& Scaramella 
1991; Blanchard, Valls-Gabaud, \& Mamon 1992; Monaco 1997(a,b); Audit, Teysser,
\& Alimi 1997; Lee \& Shandarin 1998, hereafter LS; Sheth, Mo \& Tormen 1999, 
hereafter SMT), it remains a viable model of the mass function and is widely 
used.

In this Letter, we make use of three theoretical models suggested for
the cosmological mass function:
(i) the original PS mass function $n_{PS}$ assuming the spherically 
symmetric collapse, (ii) the mass function $n_{\lambda_3}$ that
incorporates the anisotropic collapse as it is described
by the Zel'dovich approximation (LS) and (iii) the mass function 
$n_{ST}$ suggested by Sheth \& Tormen (1999, hereafter ST) and later 
derived by SMT that takes into account both the anisotropic collapse 
and some nonlocal effects. 
Recently Jenkins et al. (2001) suggested fits to mass functions obtained in
the ``Hubble Volume'' N-body simulations of some cosmological models.
We have checked that using the fits by Jenkins et al. (2001) does  
not change the conclusions of this Letter.
For comparison with observations we use mass functions obtained for cluster 
virial masses by Girardi et al. (1998) and Reiprich, B\"{o}hringer, \& 
Schuecker (2000)


Here we report the results for eight cosmological models. Among these, seven 
have recently been claimed as the best-fit models satisfying the data from 
CMB anisotropy experiments ({\sl COBE} Differential Microwave Radiometer, 
BOOMERANG-98 and  MAXIMA-1) as well as from nucleosynthesis, large-scale 
structure and type Ia SN observations. These models are labeled P11, 
BOOM+MAX+{\sl COBE}:I, BOOM+MAX+{\sl COBE}:II, PV+CMB+SN, Refined Concordance, 
MAXIMA-1, and $\Lambda$MDM. We have included the Concordance model as a 
reference model since it is often referred to as the standard $\Lambda$CDM 
model. 

None of the proponents of the best-fit models in our list have mentioned the 
explicit cluster abundance test. Rather some of them claimed that their models 
satisfy one of the many $\sigma_8-\Omega_0$ relations reported in the literature, 
others even did not apply this test at all. We have noticed more than a dozen 
predictions of the $\sigma_8-\Omega_0$ relation in the literature, some of which 
are in conflict with the others. We believe our approach here to present the 
result of the cluster abundance test is more explicit.   

This Letter is organized as follows: in $\S$ 2 we briefly
summarize the theoretical models of the cosmological mass functions,
in $\S$ 3 we briefly describe the observational mass functions,
in $\S$ 4 we outline the cosmological models, and, finally,
in $\S$ 5 we report and discuss the results.

\section{Theoretical mass functions}
The cumulative mass function (cmf)  is the comoving number density 
of gravitationally bound objects of mass greater than $M$: 
$N(>M) = \int_{M}^{\infty} \ n(M^{\prime}) dM^{\prime}$,
where $n(M) dM$ is the mass function of the collapsed objects with 
masses between $M$ and $M+dM$. The PS model based on spherical collapse of 
overdense region in a smooth  background predicts
\begin{equation}
n_{PS}(M) = F(\bar{\rho},\sigma_M) 
\ \nu \ \exp(-\frac{\nu^2}{2}), 
\label{PS}
\end{equation}
where $\nu = \delta_c/\sigma_M$, 
$F(\bar{\rho},\sigma_M) = (2/\pi)^{1/2} \ (\bar{\rho}/M^2)$ $\cdot$ 

$|d\ln\sigma_M/d\ln M|$, and  $\bar\rho$ is the
mean matter density. The canonical value $\delta_{c} = 1.686$ corresponds to 
the spherical top-hat model in the $\Omega_0 = 1$ universe. Later it was 
shown that $\delta_{c}$ only weakly depends on the background cosmology 
(\cite{eke-etal96}), and therefore we ignore it here. The rms density 
fluctuation ($\sigma_{M}$) at the mass scale M is determined by the linear 
power spectrum $\sigma_{M}^{2} = 1/2\pi^2 \hspace{0.2 cm} 
\int_{0}^{\infty} \ dk \ k^{2} \ P(k) \ W^{2}_{TH}(kR)$, 
where $W_{TH}(kR)$ is the Fourier transform of the top-hat window function.
The mass $M$ is related to $R$ as $M = 4\pi/3R^3 \bar{\rho}$. 
Although theoretically the most consistent approach requires the sharp 
k-space window function (see, e.g., \cite{bon-etal91}), we use the top-hat 
window because it results in better fit to N-body simulations (see, e.g., ST). 

The $\lambda_3$-model suggested by LS is based on the non-spherical 
collapse as described by the Zel'dovich approximation. It assumes 
that a fluid particle belongs to gravitationally bound object after it 
experiences collapse along all three principle axes. In practice it 
has been approximated by imposing the condition $\lambda_3 > \lambda_c$ 
on the smallest eigen value ($\lambda_3<\lambda_2<\lambda_1$) calculated 
for the initial density field smoothed with the sharp $k$-space 
filter corresponding to mass $M$. 
Comparisons with N-body simulations have shown that the threshold is
$\lambda_{c} = 0.37$ (Lee \& Shandarin 1999).
The mass function in this model is given as (assuming $\lambda^{\prime} 
= \lambda_c/\sigma_M$)
\begin{eqnarray}
n_{\lambda_3}(M) &=& \frac{25\sqrt{5}} {24\sqrt{2\pi}} 
\ F(\bar{\rho},\sigma_M) \ \lambda^{\prime} 
\Bigg[- 20 \lambda^{\prime} \exp\Big{(}-\frac{9\lambda^{\prime 2}}{2}\Big{)}\cr
&+& \sqrt{2\pi} \Big{(} 20\lambda^{\prime 2} - 1\Big{)}
\exp\Big{(}-\frac{5\lambda^{\prime 2}}{2}\Big{)}
{\rm erfc}\Big(\sqrt{2} \lambda^{\prime}\Big)\Bigg]\cr
&+&3\sqrt{3\pi}\exp \Big(-\frac{15\lambda^{\prime 2}}{4} \Big)
{\rm erfc} \Big( \frac{ \sqrt{3} \lambda^{\prime}}{2} \Big).   \label{LS}
\end{eqnarray}

ST suggested a correction to the PS mass function resulting in 
better fit to N-body simulations (for a discussion of motivations see SMT)  
\begin{equation}
n_{ST}(M) = F(\bar{\rho},\sigma_M) \ A \ 
v\bigg[ 1 + \bigg(\frac{\nu^2}{a}\bigg)^{q}\bigg] 
\ \nu \ \exp(-\frac{a\nu^2}{2}).
\label{ST}
\end{equation}
The parameters $A=0.322$, $a=0.707$ and $q= 0.3$, chosen by ST, have been
determined empirically from N-body simulation. 
At $A=1/2$, $a=1.0$ and $q=0$, one finds $n_{ST}= n_{PS}$. 

The cosmological parameters enter the cosmological mass function via
the shape and normalization of the linear power spectrum. 
One of the most accurate approximation of power spectrum fitting
formula incorporating baryon density wass developed by Eisenstein \& Hu 
(1998). Their formula has accuracy better then 5\% 
for baryon fraction $\Omega_b/\Omega_0$ less then 30\%. 
The cosmological models discussed here predict baryon fraction less 
then 20\%, therefore we have used the  Eisenstein \& Hu  fits for the power
spectrum. 

\section{Observational mass functions}

The predictions of the theoretical models have been tested against 
the measurements of the {\it virial} mass functions in the N-body simulations
(see, e.g., ST and references therein). Therefore, the theoretical mass 
functions must be compared with the observational virial mass functions. 

Girardi et al. (1998) provided the cumulative mass functions 
estimating the virial masses of clusters of richness $R \ge -1$
and $R \ge 1$. Both practically coincide for 
$M>10^{14.6} h^{-1}M_{\odot}$ (see Fig. 2 in Girardi et al. 1998).
This mass function is shown by filled circles in Fig. 1 and 2.
Reiprich et al. (2000) determined the cmf using X-ray flux-limited 
sample from {\sl ROSAT} All-Sky Survey. They determined the masses from
measured gas temperatures based on {\sl ASCA} observations. 
In this Letter we use the mass function corresponding
to $r_{200}$ which is usually referred to as the virial radius ({\sl open 
squares} in Fig. 1 and 2). At $M<10^{14.8} h^{-1}M_{\odot}$
the Girardi et al. mass function is significantly higher 
than that of Reiprich et al. 

It should be mentioned that the estimation of the masses is not a simple
problem. For further discussion, see, e.g., Girardi et al.(1998), 
Reiprich et al.(2000), Pierpaoli et al.(2001), and references therein.   
In addition, there is no one-to-one correspondence between optically and 
X-ray-selected clusters. There are clusters found in both optical and X-ray 
surveys, but some optical clusters do not have counterparts 
in X-ray surveys and vice versa. 
There are some evidences suggesting that the fraction of X-ray clusters
in a sample of optical clusters is smaller than the fraction
of optical clusters in a sample of X-ray clusters. If confirmed by 
further studies this means that some of optical clusters failed to become
X-ray sources by some unknown reasons.  
However, this observation has been made for 
the {\sl ROSAT} Optical X-ray Survey and 
must be taken with a great caution; it 
cannot be directly applied to any other surveys (M. Donahue 2001, private 
communication). Here we take both observational mass functions as they have 
been proposed by the authors without trying to resolve the discrepancies 
between them.   

\section{Cosmological models}

In this Letter we discuss mostly flat cosmological models that are strongly 
motivated by the inflationary model of the universe (see e.g. Ostriker \&
Steinhardt 1995 and references therein). As an illustration we have included
one open model (MAXIMA-1 with $\Omega_{tot}= 0.91$) advocated by Balbi et 
al. (2000) and one closed model ($\Lambda$MDM with $\Omega_{tot}=1.06$) 
advocated by Durrer \& Novosyadlyj (2001). Although other groups 
(Valdarnini, Kahniashvili, \& Novosyadlyj 1998 and 
\cite{pri-gro01}) have discussed $\Lambda$MDM type models, we have chosen only 
the above mentioned one for our comparison. The cosmological parameters 
have been obtained from observational data through likelihood analysis with 
various prior assumptions. 
These parameters ($\Omega_b$, $\Omega_{cdm}$,
$\Omega_\Lambda$, $n_s$, $h$, $\sigma_8$) from different models are
presented in Table 1. In our notation,
$\Omega_0 = \Omega_b + \Omega_{cdm}$, spectral index $n = n_s + n_t$.
In this letter, we have taken zero gravity wave contribution i.e.
$n_{t}=0$ with zero reionization. Among these models P11,
BOOM+MAX+{\sl COBE}:I, BOOM+MAX+{\sl COBE}:II, Concordance and MAXIMA-1 are
{\sl COBE}-normalized following the prescription of Bunn \& White (1997). For
other models we have followed the normalization suggested by the authors.
models

\section{Summary}

We have compared the theoretical predictions of cluster abundance
by several cosmological models with the observational mass functions 
determined by Girardi et al.(1998) ({\sl filled circles} in Fig.1,2) and  
Reiprich et al. (2000) ({\sl open squares} in Fig.1,2). 
In this Letter we make use of three theoretical mass 
functions $n_{PS}$, $n_{\lambda_3}$ and $n_{ST}$. It is worth 
stressing that in the range of masses ($4 \times 10^{14} h^{-1}M_{\odot} 
\le M \le 3 \times 10^{15} h^{-1}M_{\odot}$), 
the theoretical models differ one from another roughly less or similar 
to the error bars of both observational mass functions. 

At $M \le 10^{14.8} h^{-1}M_{\odot}$ no model can be reconciled 
with the Reiprich et al. (2000) X-ray mass function. 
On the other hand almost all models are in much better agreement with 
the Girardi et al. (1998) optical mass function. 
Thus, the resolution/explanation of the discrepancies between
optical and X-ray mass functions becomes crucial for the well being
of all models in question.

As far as the optical mass function is concerned the Refined Concordance, 
BOOM+MAX+{\sl COBE}:I, and  $\Lambda$MDM models show a reasonable agreement 
with observations. The P11 and especially Concordance 
models predict a slightly lower abundances than observed at 
$\sim 10^{14.6} h^{-1}M_{\odot}$. On the other hand, the 
BOOM+MAX+{\sl COBE}:II and PV+CMB+SN models predict a slightly higher
abundances than observed at $\sim 10^{14.9} h^{-1}M_{\odot}$.
The MAXIMA-1 model seems to be safely ruled out by the data on 
cluster abundances.

A similar comparison using the sharp k-space filter
for evaluation of $\sigma_M$, which is better justified for the PS
mass function (\cite{bon-etal91}), showed that all three 
theoretical mass function are systematically higher than 
that for the top-hat filter. The sharp k-space filter approach 
improves the agreement with observations for the P11 and  Concordance 
models and makes it worse for the BOOM+MAX+{\sl COBE}:II and PV+CMB+SN models. 
The conclusions for other models did not change much.

\section{Acknowledgment}
We thank referee for useful comments and acknowledge comments by 
B. Novosyadlyj. NR thanks Hume Feldman, Patrick Gorman and Surujhdeo 
Seunarine for helpful discussions and especially A. Jenkins for his help in 
developing the numerical code. We acknowledge the support of GRF 2001 grant 
at the University of Kansas.


\newpage
\centerline{Table 1}
\centerline{Parameters of the Cosmological  Models}
\begin{center}
\begin{tabular}{|c|c|c|c|c|c|c|c|}   \hline

\textcolor{red}{Models}    &\multicolumn{6}{c|}{\textcolor{red}{Parameters}}& \\ \cline{2-7}
                           &$\textcolor{blue}{\Omega_b}$
                           &$\textcolor{blue}{\Omega_{cdm}}$
                           &$\textcolor{blue}{\Omega_{\Lambda}}$
                           &$\textcolor{blue}{n_s}$
                           &$\textcolor{blue}{h}$
                           &$\textcolor{blue}{\sigma_8}$
                                                                                       
& \raisebox{3.0ex}[0pt]{\textcolor{red}{Reference}}  \\ \hline\hline
                     
\textcolor{red}{P11}
                     &0.045 &0.255  &0.7   &0.95   &0.82  &0.92  &\textcolor{red}{Lange et al.2001}
                                                                                     \\ \hline
\textcolor{green}{BOOM+MAX+{\sl COBE}:I}
                     &0.045 &0.255  &0.7   &0.975  &0.82  &0.97  &\textcolor{green}{Jaffe et al.2000}
                                                                                     \\ \hline
\textcolor{blue}{BOOM+MAX+{\sl COBE}:II}
                     &0.036 &0.314  &0.65  &0.95   &0.80  &1.06  &\textcolor{blue}{Hu et al.2001}
                                                                                     \\ \hline
\textcolor{magenta}{PV+CMB+SN}
                     &0.035 &0.245  &0.72  &1.0    &0.74  &1.17  &\textcolor{magenta}{Bridle et al.2001}
                                                                                     \\ \hline\hline\hline

\textcolor{red}{Concordance}
                     &0.03  &0.27   &0.7   &1.0    &0.68  &0.85  &\textcolor{red}{Ostriker \& Steinhardt1995}
                                                                                      \\ \hline
\textcolor{green}{Refined Concordance}
                     &0.05  &0.33   &0.62  &0.91   &0.63  &0.83  &\textcolor{green}{Tegmark et al.2001}
                                                                                       \\ \hline
\textcolor{blue}{MAXIMA-1 ($\Omega_{tot}=0.91$)}
                     &0.07  &0.61   &0.23  &1.0    &0.60  &1.05  &\textcolor{blue}{Balbi et al.2000}
                                                                                       \\ \hline
\textcolor{magenta}{$\Lambda$MDM ($\Omega_{tot}=1.06$)}
                     &0.037  &0.303  &0.69  &1.02   &0.71  &0.92 &\textcolor{magenta}{Durrer \& Novosyadlyj 2001}
                                                                                        \\ \hline  
                                                                                     
\end{tabular}
\end{center}

\newpage
\centerline{Figure Captions}

Fig. 1. Observational cmfs measured for virial mass are compared 
with different theoretical predictions: (a) P11,  
(b) BOOM+MAX+{\sl COBE}: I, (c) BOOM+MAX+{\sl COBE}: II and (d) 
PV+CMB+SN model. 
The short dash line is $n_{PS}$, long dash line is $n_{\lambda_{3}}$ 
and solid line is $n_{ST}$; the filled circles are the observational 
data points corresponding to virial masses determined by Girardi et al.
(1998) and the open squares are those determined by Reiprich et al 
(2000). The open triangle is the  value of the cmf for masses 
estimated within the $1.5 h^{-1}$ Mpc radius given by Girardi et 
al. The error bars are in 1$\sigma$ limit along the vertical direction. 
Horizontal bars indicate the bin size.

Fig. 2. Same as  Fig. 1 but with different models: (a) Concordance, 
(b) Refined Concordance, (c) MAXIMA-1 and (d) $\Lambda$MDM model.

 \newpage
 \begin{figure}
 \begin{center}
 \epsfxsize=40pc
 \epsfbox{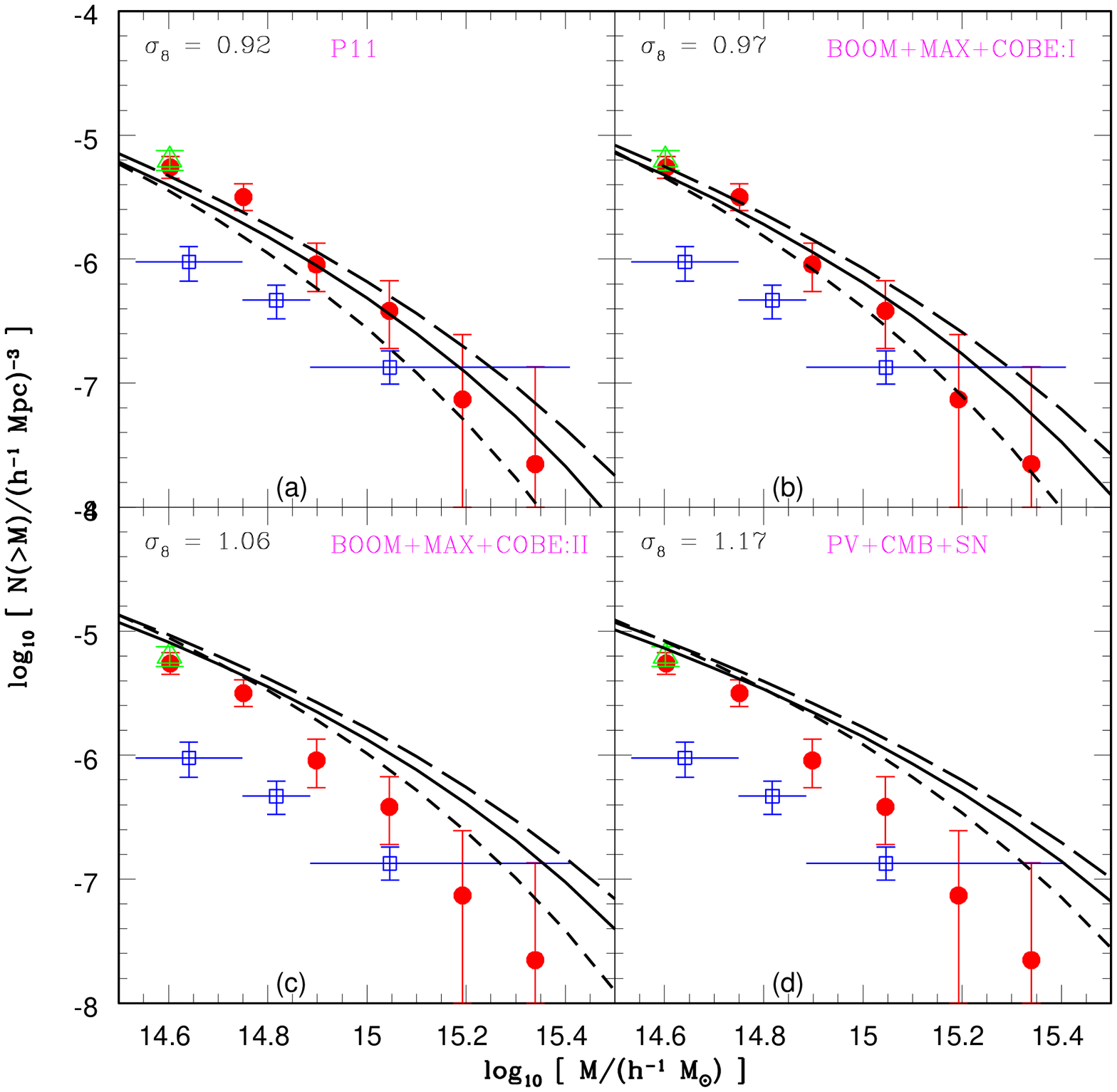}  
 \caption{} 
 \label{fig1}
 \end{center}
 \end{figure}

 \newpage
 \begin{figure}
 \begin{center} 
 \epsfxsize=40pc
 \epsfbox{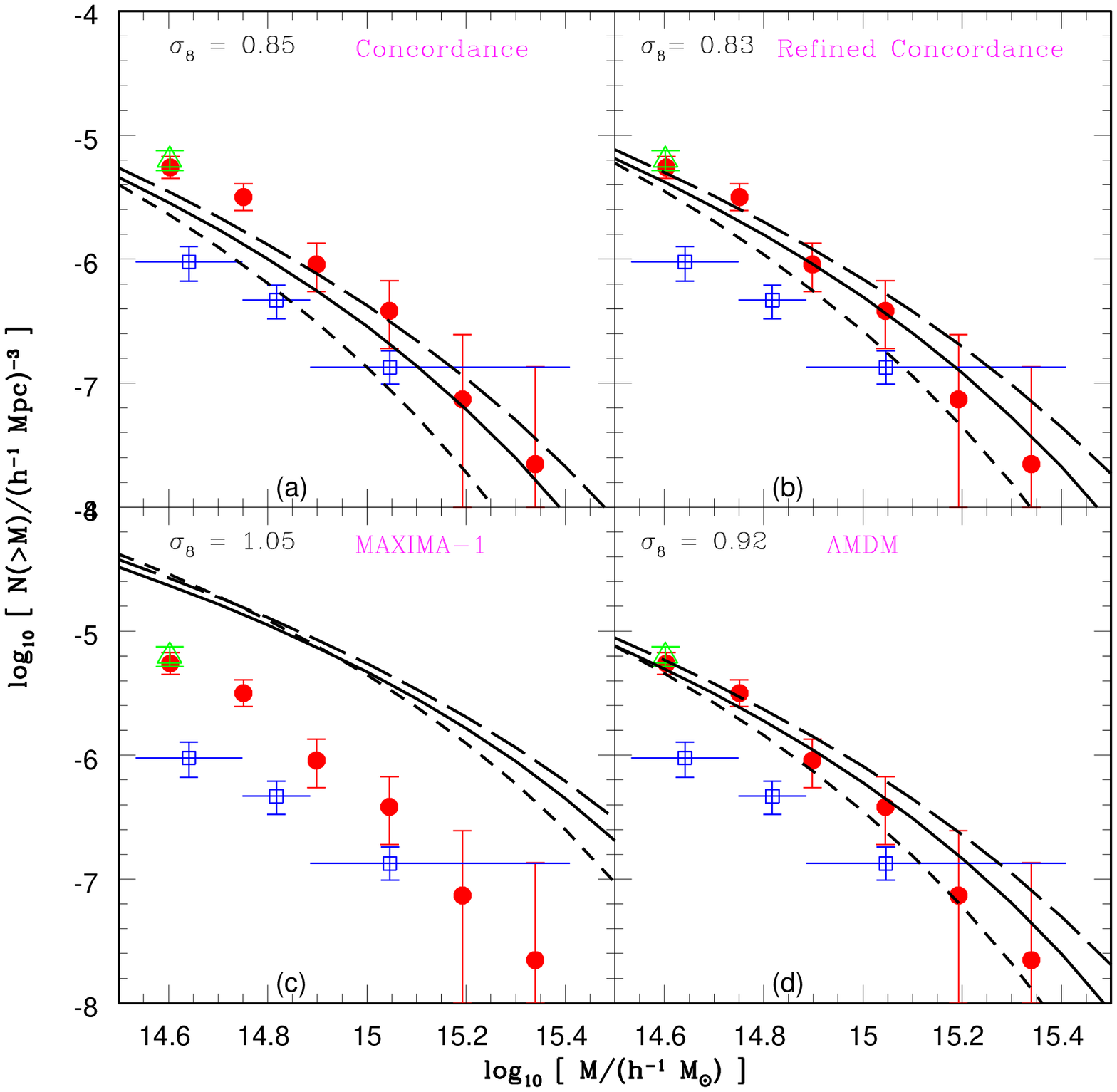}
 \caption{} 
 \label{fig2}
 \end{center}  
\end{figure}

\end{document}